\documentclass[article,twocolumn,english,footinbib,aps,prb,unsortedaddress,showpacs,superscriptaddress]{revtex4-1}
\usepackage[usenames,dvipsnames]{color}
\usepackage{amsmath}
\usepackage{amsfonts}
\usepackage{amssymb}
\usepackage{pict2e}
\usepackage{hyperref}
\usepackage{graphicx,xcolor}
\usepackage{bbding}
\usepackage{bm}

\usepackage{multirow}
\usepackage{dcolumn}  
\usepackage{float}
\usepackage{amsmath,mathrsfs}
\usepackage{amsthm}
\usepackage{epsfig}

\usepackage{array}
\usepackage{color}
\usepackage{soul}
\usepackage[caption=false]{subfig}

\newcolumntype{P}[1]{>{\centering\arraybackslash}p{#1}}

\def\Bk{{\bm k}}

\def\Vd{{\vec d}}
\def\Bg{{\bm g}}
\def\Ba{{\bm a}}

\def\Vv{{\vec v}}
\def\Vw{{\vec w}}
\def\Valpha{{\vec\alpha}}
\def\Vbeta{{\vec\beta}}
\def\Bz{{\bm z}}
\def\BR{{\bm R}}
\def\Htil{{\tilde H}}
\def\BLambda{{\bm \Lambda}}

\def\Vsigma{{\vec\sigma}}
\def\pBL{\parallel\!\BLambda\!\parallel}
\def\ie{{\it i.e.\/}}
\def\ve{\varepsilon}
\def\ns{^{\vphantom *}}
\def\sket#1{{|  \, #1 \,  \rangle}}
\def\sbra#1{{\langle \,  #1 \, |}}

\mathchardef\sOmega="710A
\mathchardef\sGamma="7100
\mathchardef\sDelta="7101
\def\Rep{\Re \,}
\def\Imp{\Im \,}

\def\frac#1#2{{\textstyle{#1 \over #2}}}
\def\half{\frac{1}{2}}

\begin{document}

\title{Band flatness optimization through complex analysis}

\author{Ching Hua Lee}
\affiliation{Institute of High Performance Computing, 138632, Singapore}
\author{Daniel P. Arovas}
\affiliation{Department of Physics, University of California at San Diego, La Jolla CA 92093, USA}
\author{Ronny Thomale}
\affiliation{Institute for Theoretical Physics, University of W\"urzburg, Am Hubland, 97074 W\"urzburg, Germany}

\date{\today}

\begin{abstract}
Narrow band electron systems are particularly likely to exhibit correlated many-body phases driven by interaction effects.  Examples include magnetic materials,
heavy fermion systems, and topological phases such as fractional quantum Hall states and their lattice-based cousins, the fractional Chern insulators (FCIs). 
Here we discuss the problem of designing models with optimal band flatness, subject to constraints on the range of electron hopping.  In particular, we show how
the {\it imaginary gap\/}, which serves as a proxy for band flatness, can be optimized by appealing to Rouch{\'e}'s theorem, a familiar result from complex analysis.
This leads to an explicit construction which we illustrate through its application to two-band FCI models with nontrivial topology (\ie\ nonzero Chern numbers).
We show how the imaginary gap perspective leads to an elegant geometric picture of how topological properties can obstruct band flatness in systems with
finite range hopping.

\end{abstract}


\maketitle

\section{Introduction}

The accumulation of electronic energy states in narrow-band systems often results in interaction-driven strongly correlated many-body phases. The limiting case in which one or more narrow bands become perfectly flat has attracted recent attention in both condensed matter and
cold atom physics \cite{tasaki,liu-arxiv}, with several proposals for realization with realistic materials 
\cite{garnica2013,n2006,suwa2010,gulacsi2010,masumoto2012,baboux2015}.  Unlike van Hove singularities \cite{PhysRev.89.1189} and other prominences
which can lead to nesting instabilities,
flat bands are featureless in momentum space, and thus tend to support similarly featureless many-body states, \ie\ without breaking of lattice symmetries. 
A classic example is the Stoner instability leading to ferromagnetism, which is rigorously established for flat band systems \cite{tasaki-prl,maksymenko2012}, but notoriously
difficult to elicit with typically dispersing bands.  In systems with attractive effective interactions, flat bands and related density profiles tend to favor a featureless $s$-wave
superconductor \cite{miya,PhysRevB.88.224512} over competing charge density wave states. Correlated states where discrete lattice symmetries are
spontaneously broken by interaction are also possible in flat band systems, especially at low filling \cite{wu-07prl070401}. In addition to broken
symmetry states, topological states known as fractional Chern insulators (FCIs) \cite{tang2011,sunfci,n2011,regnault2011a,parameswaran2013,claassen2015} have been identified in interacting nearly-flat
band models.  These systems are lattice realizations of the fractional quantum Hall effect, where their nearly flat bands effectively serve as Landau levels.

Besides the atomic limit with trivially dispersionless bands, flat bands can also arise in noninteracting systems~\cite{derzhko2015}.  Tasaki~\cite{tasaki} has
described both long-range hopping as well as local ``cell construction'' models yielding flat bands which rigorously exhibit ferromagnetism when Hubbard
interactions are included.  The class of lattices known as {\it line graphs\/}, which includes Kagome, checkerboard, and pyrochlore structures, all exhibit flat
bands at energy $E=2t$, where $t$ is the nearest neighbor hopping amplitude.  This construction was exploited by Mielke \cite{mielke-flat} to obtain
ferromagnetic ground states in the presence of a Hubbard term.  In each of these cases, the hoppings conspire to yield an extensive number of degenerate
localized modes.  Such flat band models often exhibit band touching due to additional modes from the toroidal homotopy generators~\cite{mielke-flat}.
A perfectly flat band can also result from interaction-induced self-energy renormalization~\cite{PhysRevLett.105.266403}.  For FCIs nonzero Chern number\cite{lee2015arbitrary} $C$ as well as band flatness is desired~\footnote{Note that FCIs were recently proposed for interactions scales larger than the band gap
\cite{PhysRevLett.112.126806}, as well as in a topologically trivial band~\cite{simon-arxiv}. However, nearly flat Chern bands with large band gaps are still likely most suited for stabilizing an FCI.}, though a band with $C\neq 0$ and finite range hoppings
cannot be perfectly flat \cite{seidel}. Numerical evidence indicates that this also appears true for bands with non-trivial $\mathbb{Z}_2$ topological invariants \cite{ozolins-pnas,budich-arxiv},
although this result is not yet rigorously established.

In this work, we devise a systematic approach for efficiently flattening an electronic band with a given class of hopping terms.  While any band may be trivially
flattened via band projection, \ie\ by replacing $\ve\ns_n(\Bk)\,\sket{n,\Bk}\sbra{n,\Bk}$ by $\sket{n,\Bk}\sbra{n,\Bk}$, this maneuver comes at the cost of introducing
nonlocal hoppings in real space.  Our aim here is to optimize band flatness for models with physical hoppings, which are constrained by locality.  This is achieved by
deforming a given Hamiltonian toward one with a maximal {\it imaginary gap\/} (IG).  With the help of Rouch{\'e}'s theorem from complex analysis, this problem can be reduced to one involving the analysis of a polynomial in a single variable.  To illustrate our approach, we specialize to 2-band FCIs,
and show how non-trivial band topology constrains the size of the IG, and thus the optimal band flatness. As a by-product, our mathematical approach also enables the visualization of the topological index as a certain winding number, in analogy to Volovik's interpretation of the Chern number as a winding of the Green's function in complex momentum space\cite{volovik2003universe}. 




\section{Flat bands and the imaginary gap}

Suppose we want to flatten the $m^{\rm th}$ band of a given $N$-band Hamiltonian $H(\Bk)$ with eigenvalues
$\{\ve\ns_j(\Bk)\}$. We can do so by adding a diagonal term $-\ve\ns_m(\Bk)\,\mathbb{I}\ns_{N\times N}$ to $H(\Bk)$,  so that
\footnote{This simple replacement avoids dealing directly with the band projectors which, being operators, require more mathematical care.}  
\begin{equation}
H'(\bm k)= H(\Bk)-\ve_m(\Bk)\,\mathbb{I}
\label{flattening}
\end{equation}
has eigenvalues $\{\ve\ns_1-\ve\ns_m,\ldots,0,\ldots,\ve\ns_N-\ve\ns_m\}$. To make $H'(\Bk)$ physically realistic, we perform a real-space truncation
$H'(\Bk)\rightarrow\tilde H(\Bk)$ such that $\tilde H(\Bk )$ involves only hoppings which connect unit cells separated by distances $|\BR|\le \Lambda$, where
$\BR$ is a direct lattice vector and $\Lambda$ a predefined hopping range.  The $m^{\rm th}$ band now acquires a finite bandwidth, as this truncation sacrifices
perfect flatness for the sake of finite-range hopping.

We assume that there are no generic band crossings \footnote{This is guaranteed by the Wigner - von Neumann theorem when the codimension for accidental
degeneracies is greater than the physical dimension of space, as occurs, i.e. in two-dim systems with broken time-reversal symmetry.}.
An appropriate dimensionless measure of the flatness of the $m^{\rm th}$ band is then the ratio $\Delta\ns_m/W\ns_m$ of the minimal bandgap
\begin{equation}
\Delta\ns_m=\textsf{min}\big\{\ve\ns_m(\Bk)-\ve\ns_{m-1}(\Bk)\,,\, \ve\ns_{m+1}(\Bk)-\ve\ns_m(\Bk)\big\}
\end{equation} across neighboring bands divided by the bandwidth
$W\ns_m=\textsf{max}\big\|\ve\ns_m(\Bk)-\ve\ns_{m}(\Bk')\big\|_{\Bk,\Bk'}$.  While this depends on more information than provided by the dispersion $\ve\ns_m(\Bk)$ alone, we find, for a wide variety of models studied, that the flatness is well-approximated by
\begin{equation}
f=\sum_{R>0} \ve(\BR) \Big/\sum_{R\ge\Lambda}\ve(\BR)\quad,
\label{fdecay0}
\end{equation}
where $\ve(\BR)=\int_{\hat\sOmega} {d^d\!k\over (2\pi)^d}\>\ve(\Bk)\,e^{i\Bk\cdot\BR}$ is the Fourier transform of the dispersion $\ve(\Bk)$ (dropping the band
index $m$), integrated over the first Brillouin zone ${\hat\sOmega}$.  The numerator in Eq. \ref{fdecay0} sets an overall energy scale roughly proportional to
the typical band gap \footnote{Eq. \ref{fdecay0} may not hold when certain real-space terms conspire to cancel in a special way. However, such cases are the
exception, as evidenced by the wide variety of models in Fig. \ref{fig:decay_graph} that adhere fundamentally to Eq. \ref{fdecay0}. See also Appendix A.}.
Since the bandwidth arises from the finite truncation range, it should scale approximately as the denominator.

\begin{figure}[t]
\includegraphics[scale=0.43]{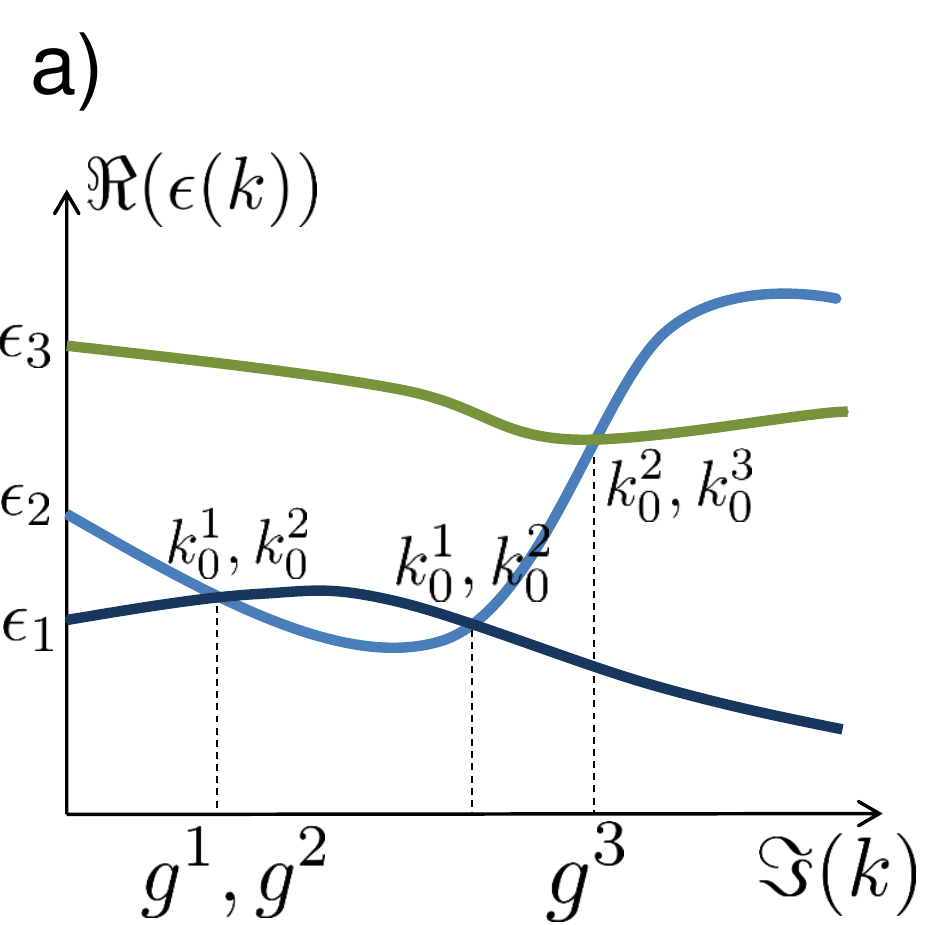}
\includegraphics[scale=0.18]{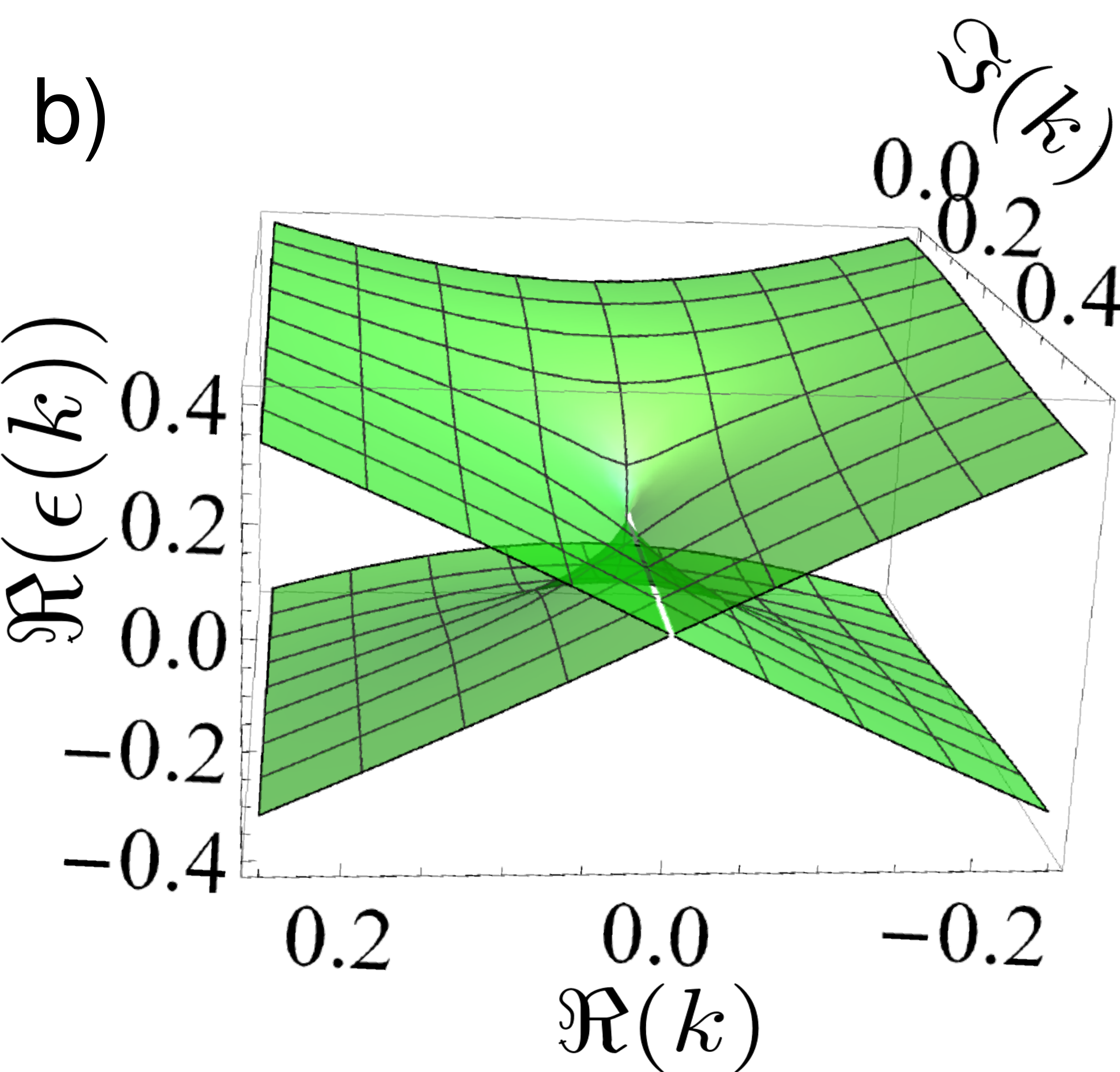}
\caption{(Color online) (a) Illustration of the band structure of a gapped Hamiltonian in the direction of imaginary momentum. The $m^{\rm th}$ band
touches another band at complex $k^{(m)}$s, with the imaginary gap (IG) $g^m=\text{min}(\Imp{k^{(m)}})$.  (b) Plot of $\Rep \ve_{1,2}(k)$
for the Dirac model $H(k)=\Vd(k)\cdot\Vsigma$ with $\Vd(k)=(\sin k,1+m-\cos k,0)$ and $m=0.3$.  The physical band gap at $\Imp{k}=0$ (back of the graph) is $2m$. The two bands, which are Riemann sheets in complex momentum space, intersect (and hence are gapless) beyond the branch point at $\Imp k\sim g\approx 0.2624$, where $H(k)$ becomes nonanalytic.}
\label{fig:riemann}
\end{figure}

\begin{figure}[t!]
\begin{minipage}{0.99\linewidth}
\centering
\renewcommand{\arraystretch}{2}
\begin{tabular}{|P{1.05cm}|P{0.8cm}|P{0.8cm}|P{1.05cm}|P{0.69cm}|P{0.69cm}|P{0.69cm}|P{0.69cm}|P{0.7cm}|}\hline
Model &\ $D0.2$ &\ $D0.4$ &\ Dwave &\ $D1$ &\ CB &\ HC &\ $D5$ &\ $D20$ \\    \hline
 $g$ &\ $0.18$ &\ $0.33$ &\ $0.79$ &\ 0.88 &\ $1.01$ &\ $1.21$ &\ $1.39$ &\ $2.94$ \\ \hline 
$f$ &\ $1.62$ &\ $3.01$ &\ $6.8$ &\ 5.82 &\ $26$ &\ $60$  &\ $37$ &\ $500$ \\ \hline  
\end{tabular}
\includegraphics[width=.99\linewidth]{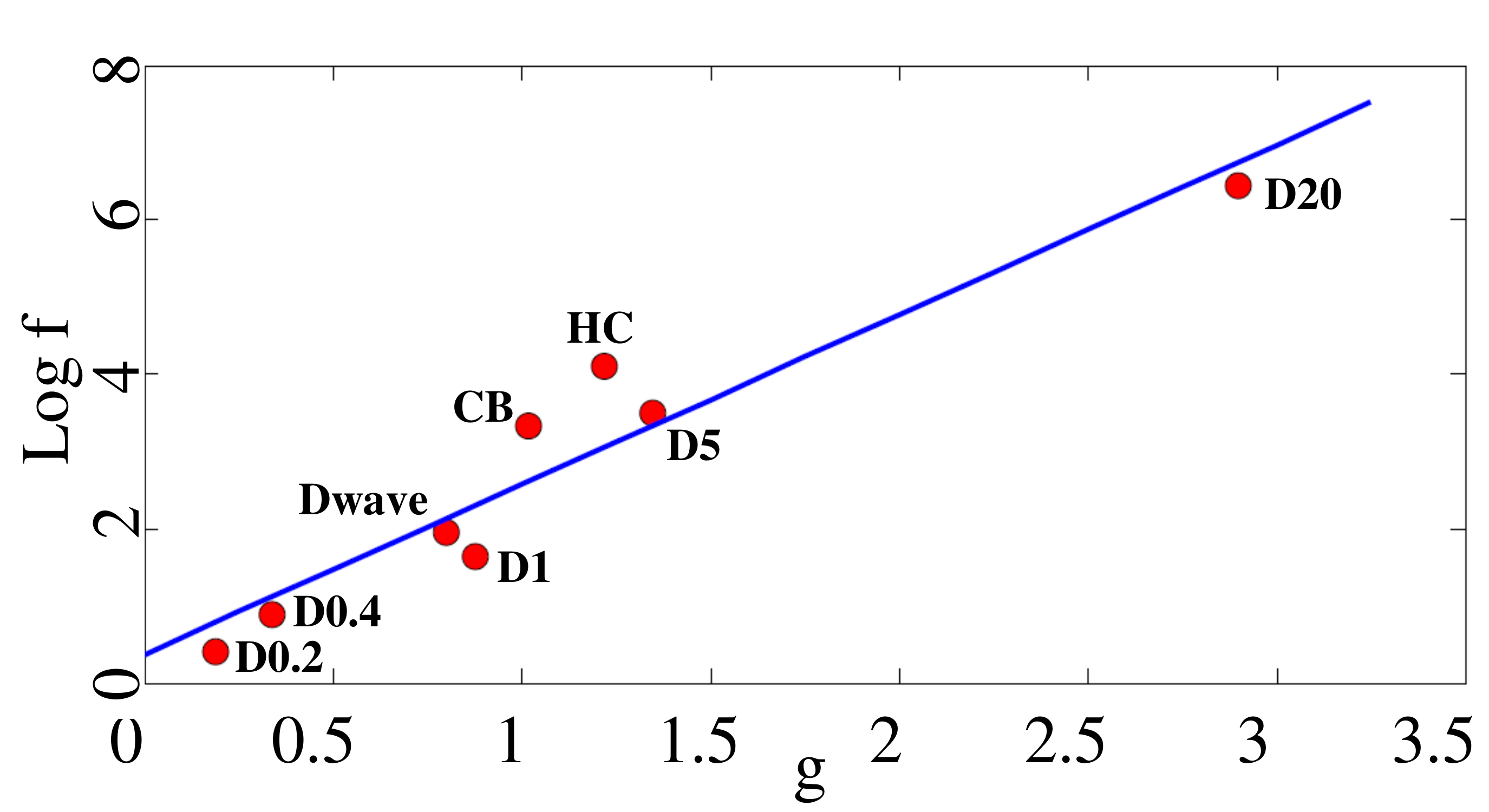}
\end{minipage}
\caption{(Color online). Imaginary gaps $g$ and flatness ratios $f$ for different 2d models flattened through Eq. \ref{flattening} and truncated to up to NNN
hoppings ($\pBL=2$).  $Dm$ refers to the Dirac model \cite{TFTTI} with mass $m$, Dwave to a $C=2$ model~\cite{lee2014},
and CB and HC to the checkerboard and honeycomb FCI models~\cite{sun2011,shengboson2011}.  $\ln f$ exhibits a strong correlation with $g$.
The linear regression coefficient of $2.22$ agrees well with $\pBL=2$, with a comparably small nonuniversal error of $r=0.11$. The largest IGs $g$ are attained by the
topologically trivial $m=5$, $20$ Dirac models, whereas HC shows the greatest $f\approx 60$ among all $C\neq 0$ models considered.}
\label{fig:decay_graph}
\end{figure}

As is well-known from classical Fourier Analysis~\cite{lang2013}, Fourier coefficients generically decay exponentially at an asymptotic rate given by the so-called
{\it imaginary gap} (IG).  The concept of the IG is employed in computing decay properties in subjects ranging from semiconductor surface science to
statistical systems and quantum entanglement~\cite{kohn1959,first1989,monch1996,he2001,thonhauser2006a,monch2014,leeandy2014,leeye2015}.  An imaginary gap
$g\ns_\mu$ can be defined for each component $k\ns_\mu$ of the wavevector as follows.  Consider the Hamiltonian $H(k\ns_\mu)$ as a function of complex
$k\ns_\mu=\Rep k\ns_\mu + i\, \Imp k\ns_\mu$, with all other wavevector components real and fixed.  Its energy manifold consists of $N$ Riemann sheets which
represent the $N$ bands $\ve\ns_n(k\ns_\mu)$.  The sheets do not touch at physical (real) wavevectors ($\Imp k\ns_\mu=0$) where $H(k\ns_\mu)$ is gapped, but one or more intersections $\ve\ns_m(k\ns_\mu)=\ve\ns_{m\pm 1}(k\ns_\mu)$ always exist at complex values of $k\ns_\mu$ known as ramification or branch points
\footnote{This occurs at the roots of the discriminant (see Appendix \ref{app:discriminant}), which always exist from the Fundamental Theorem of Algebra.}
(see Fig. \ref{fig:riemann}).  The imaginary gap $g^m_\mu$ for the $m^{\rm th}$ band is given by the magnitude of $\Imp k\ns_\mu$ minimized over all
ramification points and over all other real components of the wavevector.  Further minimizing over all directions, one obtains the overall IG $g^m=\textsf{min}\big(g^m_1,\ldots,g^m_d\big)$.  The IG $g^m$ is positive and unaffected by energy rescaling.

The Fourier transform of a given band's dispersion scales like $\ve(R)\sim e^{-gR}$ in 1-dim.  This generalizes to 
\begin{equation}
\ve(\BR)\sim\prod_{\mu=1}^d e^{-g\ns_\mu |R\ns_\mu |}
\end{equation}
in higher dims, as derived in Appendix \ref{app:multidim}), yielding a flatness parameter
\begin{equation}
f\sim\sum_{R> 0} e^{-g\ns_\mu |R\ns_\mu|} \Big/\sum_{R>\Lambda} e^{-g\ns_\mu |R\ns_\mu|}  \sim e^{\Bg\cdot\BLambda}
> e^{g \parallel\BLambda\parallel}
\label{decay}
\end{equation}
where $\Lambda\ns_\mu$ sets the maximal hopping range along the direction of elementary reciprocal lattice vector $\Ba\ns_\mu$, and 
$\pBL= \sum_\mu \Lambda\ns_\mu$ is the Manhattan distance \footnote{The decay rates from the different directions contribute
additively to the total decay exponent.}.
When $g \pBL$ is small, the inequality in Eq. \ref{decay} is far from sharp, and we expect $\ln f\approx g\,\big(\pBL+r\big)$, where $0<r<1$ is a
nonuniversal constant depending on $\BLambda$ and the $g\ns_\mu$. 

The essential insight from Eq. \ref{decay} is that a {\it maximization} of the IG $g$ leads to an {\it exponential optimization} of the flatness ratio $f$. 
Crucially, Eq. \ref{decay} extrapolates well down to small $\pBL$ despite being rigorously true only for large $\pBL$. This is empirically evidenced in
Fig. \ref{fig:decay_graph}, which shows a high correlation between $\ln f$ and $g$ for a variety of popular FCI as well as topologically trivial
models~\cite{shengboson2011,sun2011,lee2013,lee2014} with $\BLambda = (1,1)$ ($\pBL =2$), \ie\ when the above truncation procedure leaves only the nearest
and next nearest neighbor (NN and NNN) hoppings.  The value of $f$ for pre-optimized flatband models such as the checkerboard (CB) and honeycomb (HC)
models~\cite{sun2011,shengboson2011}, remains nearly unchanged after the flattening by Eq. \ref{flattening}.

In Fig.~\ref{fig:decay_graph}, largest $f$ was found for topologically trivial models. This is consistent with the fact that a model with $C\neq 0$ and finite hopping range
can never be completely flat.  This was proven in Ref.~\onlinecite{seidel} via K-theory, where it was shown that the band projectors of such models must be Laurent Polynomials with finite order in $z\ns_\mu\equiv e^{ik\ns_\mu}$. Consequently, complex singularities must be present which, in our context, imply inevitable truncation effects
resulting in a nonzero bandwidth. The converse is more subtle: A perfectly flat band with $f=0$ ($g=\infty$) can still be topologically nontrivial ($C\neq 0$) if the hoppings are not truncated. Examples include lattice models with Gaussian hoppings in a magnetic field \cite{kapit2010}, which are inspired by parent Hamiltonians~\cite{PhysRevLett.99.097202,PhysRevB.80.104406,rashba1997} for the chiral spin
liquid. 
Another subtlety is that chiral edge states, which are usually associated with a nontrivial Chern number, can occur in Floquet systems with flat bands
\cite{rudner2013,fulga2015}. 

\section{Analytical optimization of the imaginary gap} 

The imaginary gap (IG) $g$ of $H(\Bk)$ provides good lower-bound of the flatness ratio $f\gtrsim e^{g\pBL}$ for the bands of the flattened $\tilde H(\Bk)$. The next step is to compute $g$ efficiently. 
We want to generate a local $N\times N$ model $\tilde H(\Bk)$ with an almost flat band and hopping terms satisfying $\tilde H(\BR)=0$ for
$|R\ns_\mu| > \Lambda\ns_\mu$. This can be done via Eq. \ref{flattening}, followed by a truncation of the resulting $H(\Bk)\rightarrow \Htil(\Bk)$.
Expressed in terms of the $z\ns_\mu=e^{ik\ns_\mu}$, we have $\Htil_{ij}(\Bz)=\big[\Htil_{ji}(\Bz)\big]^*$, with each $\Htil\ns_{ij}(\Bz)$ a Laurent polynomial in each of the
$z\ns_\mu$ with powers ranging from $z_\mu^{-\Lambda_\mu}$ to $z_\mu^{+\Lambda_\mu}$; note that $z^{-1}_\mu=z^*_\mu$ for $z$ on the unit circle, the boundary of the analytic continuation region.  The energy eigenvalues
$\ve\ns_n(\Bz)$ are roots of the characteristic polynomial
\begin{equation}
P(\ve;\Bz)=\textsf{det}\big[\ve\,\mathbb{I}\ns_{N\times N} - \Htil(\Bz)\big]\ .
\label{charpoly}
\end{equation}
The energy manifold is singular at roots of the discriminant,
\begin{equation}
D(\Bz)= \prod_{m<n}^N\big[\ve\ns_m(\Bz)-\ve\ns_n(\Bz)\big]^2\ ,
\label{discriminant}
\end{equation} 
which is defined for any $N$.  As shown in Appendix \ref{app:discriminant}, $D(\Bz)$ can be expressed~\cite{brooks2006} in terms of the coefficients $p\ns_l(\Bz)$ of $P(\ve;\Bz)=\sum_{l=0}^N p\ns_l(\Bz)\,\ve^{N-l}$, with $p\ns_0\equiv 1$.  For $\Bk\in\mathbb{R}^d$, the coefficients are real, because they are symmetric polynomials in the eigenenergies: $p\ns_1(\Bz)=-\sum_j \ve\ns_j(\Bz)$, $p\ns_2(\Bz)=\sum_{j<l} \ve\ns_j(\Bz)\,\ve\ns_l(\Bz)$, {\it etc.\/}
From Eq.~\ref{charpoly}, each $p\ns_l(\Bz)$ is a polynomial in each $z\ns_\mu$ with negative degree $-l\Lambda\ns_\mu$ and positive degree $+l\Lambda\ns_\mu$.
In what follows, it suffices to know know that $D(\Bz)$ is itself a multinomial of maximal degrees $\pm M_\mu=\pm N(N-1)\Lambda\ns_\mu$ in each $z\ns_\mu$.
For local models $N=2$ band models which can be written as $\Htil(\Bz)=\Vd(\Bz)\cdot\Vsigma$ \footnote{Throughout, we use boldface to denote vectors in position/momentum space, and
arrows to denote vectors in internal (spin) space.}, the discriminant reduces to the familiar expression $D(\Bz)=\sum_{j=1}^3 d_j^2(\Bz)$.

We are now ready to optimize the IG. We first compute, in each direction $\mu$, the IG $g\ns_\mu=\textsf{min}\, |\Rep \ln \xi\ns_\mu|$,
where $\xi\ns_\mu$ is a root of $D(\ldots,z\ns_\mu,\ldots)$, with all $z\ns_{\mu'}$ for $\mu'\ne\mu$ considered as parameters with respect to which the
minimization is performed.  Expressed as a Laurent polynomial, the discriminant may be written as
\begin{equation}
D(z)= \sum_{n=-M}^{M} \!\! D_n z^n \ ,
\label{deteq}
\end{equation}
where $D\ns_{-n}=D_n^*$, and where we have dropped the direction index $\mu$.  We now analytically continue to\footnote{The analytic continuation of
a function onto a domain is uniquely defined by its boundary values. Otherwise, the difference between two different continuations will be trivially zero on the
boundary but nontrivial within the domain. Here, $z^*_j=z_j^{-1}$ on the boundary $|z_j|=1$, which uniquely defines the analytic continuation
shown.}  $|z|\ne 1$.  The Hermiticity of $\Htil$ guarantees that if $D(z)=0$, then $D(1/z^*)=0$, hence \emph{exactly} $M$ of the $2M$ roots of the analytic function $z^M D(z)$ will lie within
the unit circle $|z|=1$. The IG is then determined by the root lying closest to $|z|=1$. 

The task of finding this root is greatly facilitated by Rouch{\'e}'s
theorem~\cite{lang2013}, which states that if $\big|f(z)\big| > \big| h(z) - f(z) \big|$ on a closed contour ${\cal C}$, then $f(z)$ and $h(z)$ have the same number of
zeros within ${\cal C}$. To understand this intuitively, consider a man at $f(z)$ walking a dog at $h(z)$ near a tree. Let $\textsf{arg}(z)$ denote the winding around the tree. If the dog's leash is shorter than the minimal distance of the man from the tree, the dog and the man must encircle the tree the same number of times.

Now let $f(z)=D\ns_0\, z^M$ and $h(z)=z^M D(z)$, with the contour ${\cal C}$ being the circle $|z|=\sDelta$.  Clearly $f(z)$ has an $M$-fold degenerate root at $z=0$
and no others.  Since $\big|\sum_{n\ne 0} D\ns_n\,z^n \big| < \sum_{n\ne 0} |D\ns_n|\,\sDelta^{M+n}$ on ${\cal C}$, where the sums are over $n\in\{\pm1,\ldots,\pm M\}$,
Rouch{\'e}'s theorem then guarantees that if $|D\ns_0| \sDelta^M > \sum_{n\ne 0} |D\ns_n| \, \sDelta^{M+n}$, the function $h(z)=z^M D(z)$ also has $M$ roots within
${\cal C}$. Since the rhs of the inequality increases without bound for $\Delta\gg 1$, we conclude that $g\ns_\mu > -\ln\Delta$, where $\Delta$ is the smallest
positive root of
\begin{equation}
F(\Delta)\equiv\sum_{n=1}^M |D\ns_n|\big(\Delta^{M+n} + \Delta^{M-n}\big) - |D\ns_0|\,\Delta^M \ .
\label{rouche}
\end{equation}
The problem of finding a lower bound for the IG has been reduced to the simpler problem of solving a real polynomial equation $F(\Delta)=0$.  Essentially, we sacrificed an exact determination of $g\ns_\mu$ to settle for a lower bound, and at the same time avoided the necessary step of finding the arguments of the roots of $h(z)$.  We shall see below that this lower bound is already sufficient in providing an estimate of $f$. 

Eq.~\ref{rouche} can be solved numerically, and in certain cases analytically via the substitution
\begin{equation}
U=\Delta+\Delta^{-1}.
\end{equation}
An optimally flat model may be obtained
by varying $H(\Bk)$ until the root $\Delta>0$ in Eq.~\ref{rouche} is minimized. If $\Delta < 1$ is maintained throughout the minimization, no branch point ever touches the unit circle, \ie\ the physical gap never closes, and we remain in the same topological class.

\section{Two-band Chern models}

Many of the important flat band models such as the Honeycomb, Checkerboard and Dirac models contain only $N=2$ bands and NN hoppings
($\pBL=1$).  Their discriminants are at most of quadratic ($M=2$) degree, and can be readily studied and optimized analytically. 
We write the truncated Hamiltonian as $\Htil(z)=\Vd(z)\cdot\Vsigma$ where $z=e^{ik}$ for a given momentum component.  The $\Vd$ vector takes the form 
\begin{equation}
\Vd = 2 (\Vw \cos k - \Vv \sin k)+ \Vbeta\quad ,
\label{dvector}
\end{equation}
where $\Vw$, $\Vv$, and $\Vbeta$ are real 3-component vectors that depend parametrically on the other momenta.
With $\Valpha\equiv\Vw+i\Vv$, we can rewrite $\Vd$ as $\Vd= \Valpha\,z + \Valpha^{\,*}\,z^{-1} + \Vbeta$.  The coefficients in the discriminant may now be read off:
$D\ns_0=2\Valpha\cdot\Valpha^{\,*} + \Vbeta\cdot\Vbeta$, $D\ns_1=2\Valpha\cdot\Vbeta$, and $D\ns_2=\Valpha\cdot\Valpha$.
Substituting these expressions in Eq. \ref{rouche} and letting $U =\Delta+\Delta^{-1}$ , we obtain 
\begin{equation}
U=\sqrt{\left|{D\ns_1\over 2 D\ns_2}\right|^2+\left| {D\ns_0\over D\ns_2}\right|}-\left|{D\ns_1\over 2 D\ns_2}\right|\ .
\label{pq}
\end{equation}
Since $\Delta\leq 1$, we choose the root $\Delta=\half\big(U-\sqrt{U^2-4}\,\big)$, yielding a lower bound for the overall flatness ratio $f$ in the direction $\mu$:
\begin{equation}
f >\mathop{ \textsf{min}}_\mu \bigg\{\half U\ns_\mu + \sqrt{\frac{1}{4}U_\mu^2-1}\bigg\}\ ,
\label{r}
\end{equation}
which is monotonically increasing in $U_\mu$, where $\Delta\ns_\mu$ and $U\ns_\mu$ are the \emph{minimal} $U$ and \emph{maximal} $\Delta$ in direction $\mu$,
optimized over the wavevector components in all other directions. Note that the flatness ratio $f$ increases with decreasing $|D\ns_2|$ when $U$ is
sufficiently large.  In terms of the original 3-vectors, 
\begin{equation}
\begin{split}
|D_0| &= 2|\Vw\,|^2+2|\Vv\,|^2+|\Vbeta\,|^2 \\
|D_1| &= 2\sqrt{\big(\Vw\cdot\Vbeta\big)^2 + \big(\Vv\cdot\Vbeta\big)^2}\\
|D\ns_2|&=\sqrt{ \big(|\Vw\,|^2-|\Vv\,|^2\big)^2 + 4\big(\Vw\cdot\Vv\big)^2}\ ,
\end{split}
\end{equation}
which are rotationally invariant, consistent with the basis independence of $\Htil(z)$.  Note also from Eq.~\ref{pq} that $f$ is unaffected by an overall rescaling
of $\Vw$, $\Vv$, and $\Vbeta$.  To maximize $U$, and hence $f$, we want $|D\ns_{1,2}| \ll |D\ns_0|$.

\subsection{Topological constraint on flatness} 

\subsubsection{$D_2=0$ cases}

The parametrization in terms of $\vec w,\vec v$ and $\vec \beta$ suggests a geometric interpretation. Various FCI models belong to the simplest case of
$D\ns_2=0$, where $|\Vw\,|=|\Vv\,|$ and $\Vw\cdot\Vv=0$; for $D\ns_2\neq 0$  cases see the next subsection. From Eq. \ref{pq}, 
\begin{eqnarray}
U=\left|{D\ns_0\over D\ns_1}\right|&=&{|\Vw\,|^2+|\Vv\,|^2+\half |\Vbeta\,|^2\over |\Vw\,|\,|\Vbeta\ns_\parallel|}\ ,
\end{eqnarray} 
where $\Vbeta\ns_\parallel$ is the component of $\Vbeta$ in the plane spanned by $\Vw$ and $\Vv\,$: $\Vbeta\equiv\Vbeta\ns_\parallel + \Vbeta\ns_\perp$.
To optimize flatness, $\Vbeta$ must avoid the largest possible torus of constant $U$, defined by
\begin{equation}
|\Vbeta\ns_\perp|^2 + \big(|\Vbeta\ns_\parallel\,|- U\,|\Vw\,|\big)^2 = |\Vw\,|^2 \big(U^2-4\big).
\label{big}
\end{equation}
For large $U$, its approximate inner and outer radii are
$2 U^{-1} |\Vw\,|$ and $2U |\Vw\,|$.  Thus $\Vbeta$ should either have a small magnitude inside the `donut hole', or a large one outside the torus.

\begin{widetext}
\begin{figure*}
\includegraphics[width=.9\linewidth]{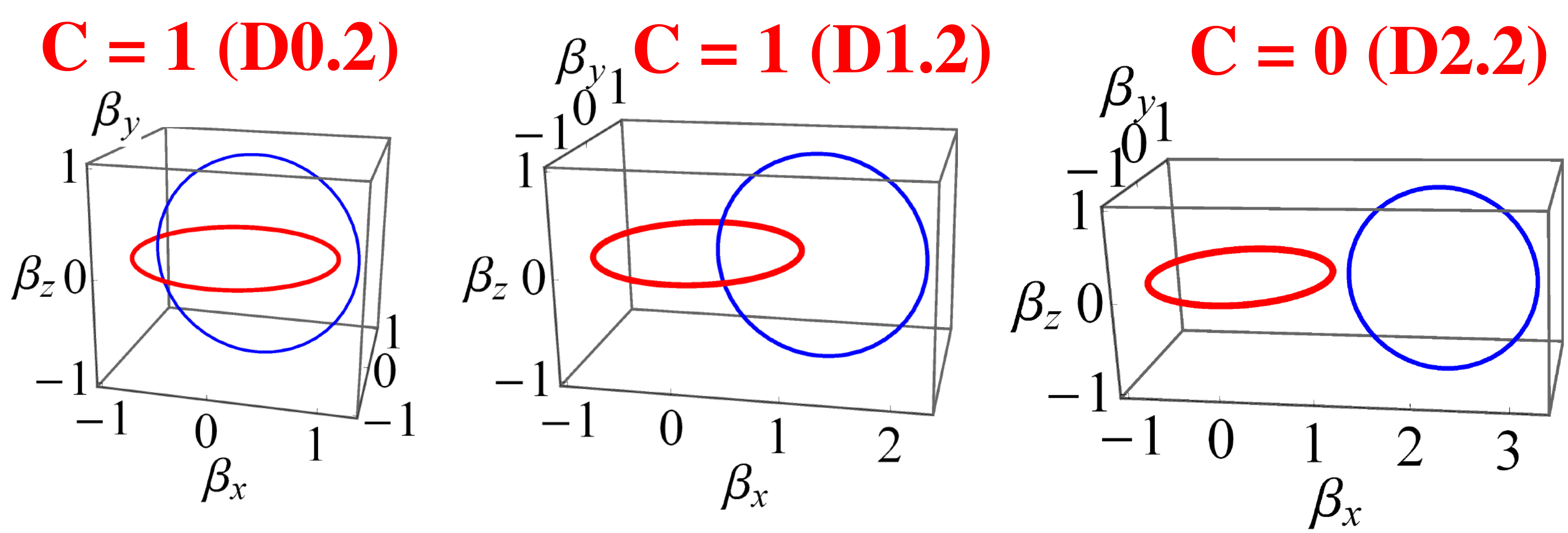}
\caption{(Color online). Nodal ring $U=U_x=2$ representing gap closure (red, Eq. \ref{pq}), and $\Vbeta$ loop (blue, Eq. \ref{dvector}), representing the $\Vd$
vector of Eq. \eqref{dirac} for $m=0.2$, $1.2$, and $2.2$. The linking number between the ring and the loop is $1$ in the $m<2$ regime where $C=1$,
and zero otherwise. The two loops are furthest separated at $m=1$, which also makes that the case with highest flatness ratio (see Eq. \ref{big}).}
\label{fig:windings}
\end{figure*}
\end{widetext}

Consider optimizing $f$ in the $x$-direction for a 2-dim model, so that $\Vbeta=\Vbeta(k\ns_y)$ traces out a loop as $k\ns_y$ varies over a period.
To remain in the same topological class, $\Vbeta$ must not pass through any point where the gap closes, \ie\ where $U\ns_x=2$, which occurs when $|\vec \beta|=|2\vec w|$. This is just the ring of radius $|2\vec w|$ in the plane spanned by $\vec v$ and $\vec w$, centered at its origin
(Fig.~\ref{fig:windings}). Configurations belonging to the same topological class thus are those that can be reached without intersecting this nodal ring, \ie\ 
those $\Vbeta$ loops have the same {\it winding number} around the ring.  To maximize $\textsf{min}(U\ns_x)$, we can either increase the size of the loop
$\Vbeta(k\ns_y)$, or shift it far away from the origin.  A large loop, however, entails large coefficients of the terms in $k\ns_y$, which will lead to small $U\ns_y$
when the same procedure is applied to the $k\ns_y$ direction. Hence a model with minimal $f$ in both directions should have loops $\Vbeta(k\ns_x)$ and
$\Vbeta(k\ns_y)$ of radii of the same order of magnitude as $|\Vw\,|=|\Vv\,|$.

It is now clear how topology constrains $f$: In the topologically trivial case, $\Vbeta$ need not wind around the nodal ring, yet can still entail arbitrarily large
$\textsf{min}(U\ns_x)$ and $\textsf{min}(U\ns_y)$ by being far from the ring.  By contrast, non-trivial topology requires that the $\Vbeta$ loop winds around
the nodal ring, constraining its size and position. 

\subsubsection{Example: 2d Dirac model}
Consider the 2d Dirac hamiltonian 
\begin{equation}
\Htil(\Bk)=\sin k\ns_x\, \sigma^x + \sin k\ns_y \, \sigma^y + (m+\cos k\ns_x +\cos k\ns_y)\,\sigma^z\ ,
\label{dirac}
\end{equation}
which is Eq. \ref{dvector} with $k\to k\ns_x$, $\Vw=(0,0,\half), \Vv=(-\half,0,0)$, and $\Vbeta = (0, \sin k\ns_y,m+\cos k\ns_y)$.  We have $|D\ns_2|=0$,
$ |D\ns_1|=m+\cos k\ns_y $ and $|D_0|=2+2m\cos k\ns_y +m^2$. Eq.~\ref{dirac} is symmetric in $k_x$ and $k_y$, so $g_x=g_y$ and we only need to consider one
direction, $k\ns_x$. The ratio that determines the band flatness is $U\ns_x=|D\ns_0/D\ns_1|$.  It attains extremal values when $\cos k\ns_y = \pm 1$, where 
\begin{equation}
U_x=\left|{D\ns_0\over D\ns_1}\right|=\left|\, m\pm 1 + {1\over m\pm 1\,}\right| \ .
\label{diracbounds}
\end{equation}       
We then obtain the flatness ratio $f\geq\Delta_x^{-2}=\Delta_y^{-2}$ by choosing the larger of the solutions to $\Delta\ns_x=\Delta=\half\big(U-\sqrt{U^2-4}\,\big)$. The optimal flatness ratio bound is obtained at $m=\sqrt{2}$, where $f=3+\sqrt{8}\approx 5.82$. In this case, the bound set by Rouch\'{e}'s theorem is saturated.  A numerical computation from
$\Htil(\Bk)$ gives an actual flatness ratio of $f\approx 6$, which is close to our lower bound.
One can verify that, in this case, the inequality in Rouch\'{e}'s theorem is saturated for {\it all\/} values of $m$.
Geometrically, we see that $\Vbeta$ describes a circle of radius unity: $(\beta\ns_y-m)^2+\beta_x^2=1 $.  As shown in Fig.~\ref{fig:windings}, it has a
linking number of $1$ with the nodal circle $\beta_x^2+\beta_y^2=1$ for $0<|m|<2$, \ie\ $C=\pm 1$.



\subsubsection{General cases with nonzero $D_2$}
\label{genericD2}

\begin{figure}
\centering
\includegraphics[scale=0.45]{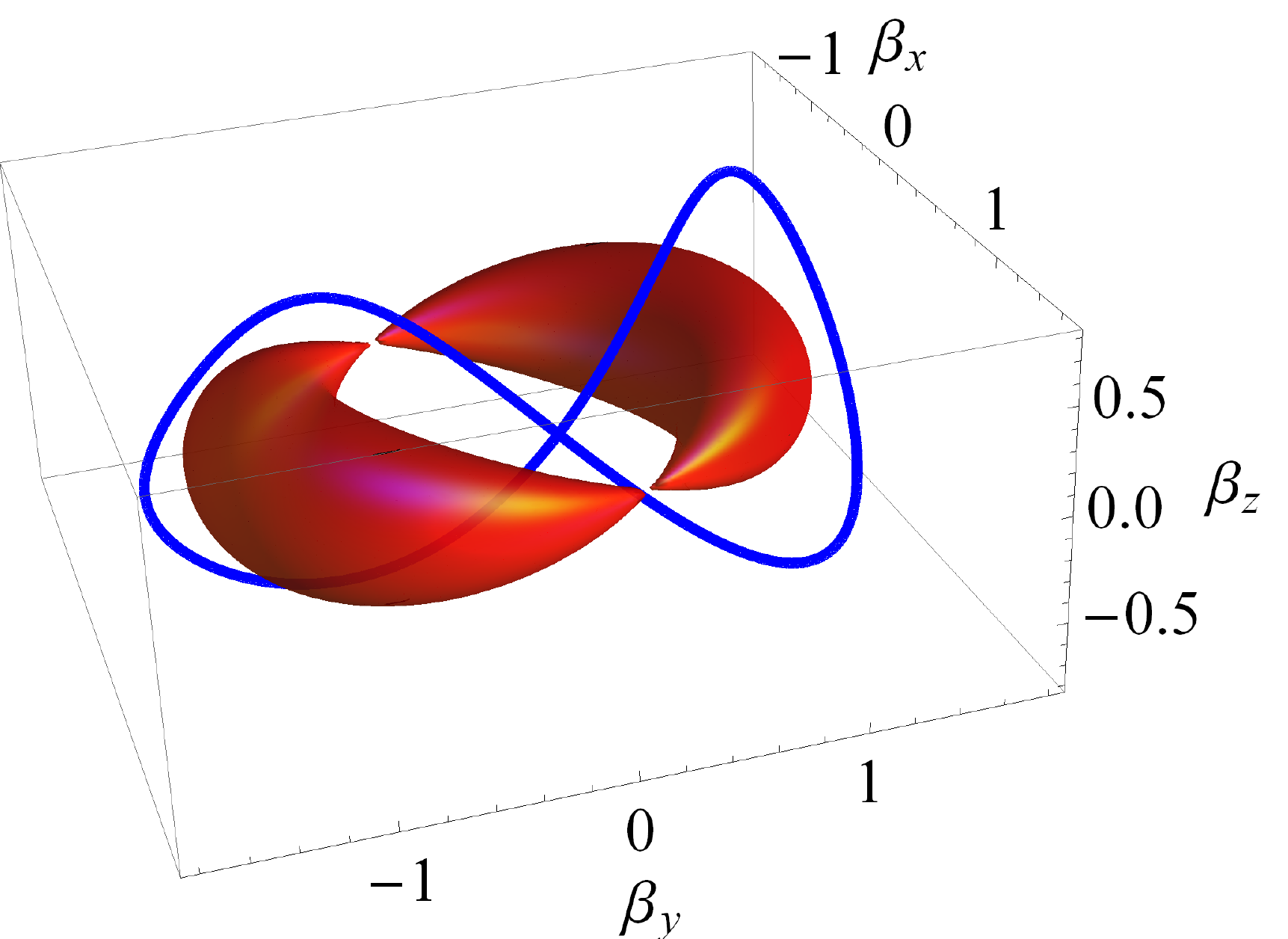}
\caption{(Color online) An illustration of the nodal surface (red) and $\vec\beta$ loop (blue) for the general case $D_2\neq 0$, with constant $|v|=1.05\,|w|$ and
$\Vw\cdot \Vv=0.05\,|\vec w||\vec v|$. Shown is  $\Vbeta$ with linking number $2$.}
\label{fig:windings2}
\end{figure}

We now discuss the geometric picture for general two-dimensional $2$-band models with $D\ns_2$ not necessarily zero. From the general expression of $U$ in Eq. 9 of the main text, we find that the nodal
points (where $U=2$) occur at $|D\ns_0|=\half |D\ns_1|+\half |D\ns_2|$. As shown in Fig. \ref{fig:windings2}, the nodal ring in general broadens to become two
bean-shaped surfaces that intersect at two points. In general, their exact shape will also depend on the other momentum parameters.
 
Most importantly, this general case is topologically identical to the $D\ns_2=0$ case. The Chern number remains the winding number of $\Vbeta$ around
the nodal region, which still has the same topology as the ring, except that there are two additional topologically trivial regions inside each of the bean-shaped
surfaces. $\Vbeta$ loops inside them are limited to small values of $U$, and are of limited usefulness to the search of flatband models with large $f$.

\section{Conclusion}

We have restated the band flattening problem for a truncated Hamiltonian $\Htil(\Bk)$ with finite range hoppings in terms of the optimization
of the {\it imaginary gap\/}, which is the smallest imaginary component of the wavevector for which the Hamiltonian $\Htil(\Bk)$ is singular.  Appealing to Rouch{\'e}'s
theorem, this optimization is further reduced to an analysis of a finite order polynomial, and finally to a vector geometry problem. Our approach provides geometric insight
on how a nonzero Chern number imposes a finite bandwidth for short-range hopping models. It also offers a constructive approach to optimizing the band flatness
of short-range hopping models.



R. Thomale thanks J.~Budich and M.~Maksymenko for discussions. 
R. Thomale is funded by the European Research Council through ERC-StG-TOPOLECTRICS-336012 and by DFG-SFB 1170. 

\bibliography{refs-flat,main}


\appendix

\section{Decay properties of the eigen-energies $\ve(k_x,k_y)$ in real-space and the imaginary gap}
\label{app:multidim}

We provide a derivation that the scaling behavior of real-space hoppings $\ve(\BR)$ is given by  $\prod_{\mu=1}^d e^{-g\ns_\mu |R\ns_\mu|}$, where $g\ns_\mu$
is the imaginary gap for $\Bk$ parallel to the elementary reciprocal lattice vector ${\bm b}\ns_\mu$, with other components of $\Bk$ held fixed.  This result forms
the basis of Eq. 3 of the main text. For ease of notation, we  specialize to the case of two dimensions $\BR=(X,Y)$. 
First, we clarify how the analytic continuation is performed.  The energy of a particular band $\ve(k\ns_x,k\ns_y)$ is a function of two variables,
which we analytically continue to the complex plane one at a time, while regarding the other as a parameter, \ie\ $\ve(k\ns_x,k\ns_y)\to \ve(z,k\ns_y)$ with
$z=e^{ik\ns_x}$.

The Fourier decay rate $g\ns_x(k\ns_y)$ can be found by finding the location of the singularity of $\ve(z,k_y)$ closest to the unit circle $|z|=1$.
Analyzing $\ve(k\ns_x,z)$ with $z\equiv e^{ik\ns_y}$ yields $g\ns_y(k\ns_x)$.
We now find the asymptotic bound on $\ve(X,Y)$. First, we Fourier transform over $k\ns_x$: 
\begin{equation}
\begin{split}
\ve(X,Y)&=\int\limits_{\hat\sOmega}{d^2\!k\over (2\pi)^2}\>\ve(k\ns_x,k\ns_y)\,e^{i(k\ns_x X + k\ns_y Y)} \\ 
&=\int {d k\ns_y\over 2\pi}\>\ve(X,k\ns_y)\,e^{ik\ns_y Y} \\
&\sim\int {d k\ns_y\over 2\pi}\> e^{-g\ns_x(k\ns_y) |X|}\,e^{i\theta(X,k\ns_y)}\,e^{ik\ns_y Y}\ ,
\end{split}
\end{equation}
where we have invoked $|\ve(X,k_y)|\sim e^{-g\ns_x(k\ns_y)|X|}$ \cite{lang2013}.  The quantity $\theta(X,k\ns_y)$ represents an unknown phase that turns
out to be irrelevant.  Next we do the $k\ns_y$ Fourier transform.  We obtain a simple bound upon expanding about the minimum $g\ns_x$ of $g\ns_x(k\ns_y)$
\begin{equation}
\begin{split}
\ve(X,Y)&\sim\Bigg| \int{dk\ns_y\over 2\pi}\>e^{-g\ns_x(k\ns_y) |X|} \,e^{i\theta(X,k\ns_y)}\,e^{ik\ns_y Y}\Bigg| \\
&\le  \int{dk\ns_y\over 2\pi}\> \Big| e^{-g\ns_x(k\ns_y) |X|} \Big| \\
&= \int{dk\ns_y\over 2\pi}\> e^{-g\ns_x |X|}\,e^{-g''_x(k\ns_y-k^0_y)^2 |X|/2+\ldots}\\
&\approx \big(2\pi |X|\, g''_x\big)^{-1/2} e^{-g\ns_x |X|} \sim e^{-g\ns_x |X|}\ ,
\end{split}
\end{equation}
where $k^0_y$ is the value of $k\ns_y$ where $g\ns_x(k\ns_y)=g\ns_x$ is minimized, and $g''_x$ is the curvature at that point. The above approximation is
justified in the limit of large $|X|$, where higher-order terms in $(k\ns_y-k^0)$ are rapidly suppressed.  As such, only contributions from $g\ns_x(k\ns_y)=g\ns_x$
and a small neighborhood around it are non-negligible. Note that we have replaced the periodic integral over $k\ns_y$ with an infinite integral above,
so the former will not be strictly correct in the limit of constant $g\ns_x(k\ns_y)$. Still, the result $\ve(X,Y)\sim e ^{-g\ns_x |X|}$ holds in that case.

If we repeat the above derivations starting from the partial Fourier Transform $\ve(k\ns_x,Y)$ instead, we obtain an analogous bound involving $g\ns_y$. Combining these results, we obtain
\begin{equation}
\ve(X,Y)\sim e^{-g\ns_x |X| - g\ns_y |Y|} < e^{-g\, \parallel \BR \parallel}\ .
\label{decayapp}
\end{equation}
Eq. 3 of the main text predicts a flatness ratio of $f\sim e^{g \pBL}$ after a real-space truncation of $\ve(X,Y)$ that retains only terms within $|X| \leq \Lambda\ns_x$ and $|Y|\leq \Lambda\ns_y$. This ratio depends crucially on Eq. \ref{decayapp}, which is exact only in the asymptotic limit of large $\Lambda$. In practice,
however, it provides excellent agreement with numerical results even for $\pBL =||(1,1)||=2$, as shown explicitly in Fig. 2 of the main text,
and in the example on the Dirac Model (also see main text). 
As mentioned, $g\ns_\mu$ only rigorously controls the real-space decay rate asymptotically. Furthermore, the derivation leading to Eq. \ref{decayapp} also contains large $|\BR|$ approximations. There may also be certain peculiarities in the shape of $\ve$ that suppresses certain Fourier components, {\it e.g.\/}, the case of the
D-wave model, which has a poor overlap with the first harmonics $\cos k\ns_x$ and $\cos k\ns_y$. These will lead to an anomalous decay not captured in the
asymptotics. When $g$ is small, the next smallest truncated terms will not be much smaller than the leading truncated terms, being only suppressed by a factor
$e^{-g}$, and the decay rate should in fact lie between $g\pBL$ and $g\big(\pBL+1\big)$.

\section{More on the discriminant}
\label{app:discriminant}
\subsection{Explicit form}
The discriminant of a polynomial
\begin{equation}
P(\ve;\Bz)=\sum_{l=0}^{N} p\ns_l(\Bz)\,\ve^{N-l}\ ,
\end{equation}
with $p_0=1$, can be expressed in terms of the resultant of $P(\ve)$ and its derivative $P'(\ve)$ (with $\Bz$ suppressed). The resultant is proportional to the
determinant of the $(2N-1)\times(2N-1)$ Sylvester matrix shown below, where the first $N-1$ rows consists of the coefficients of $P(\ve)$ and the next $N$ rows the
coefficients of $P'(\ve)$. Written out explicitly, the discriminant is equal to $(-1)^{N(N-1)/2}/p\ns_N$ times the determinant of the Sylvester matrix
\begin{widetext}
\begin{equation}
\left[\begin{matrix}
 & p\ns_N & p\ns_{N-1} & p\ns_{N-2} & \ldots & p\ns_1 & p\ns_0 & 0 \ldots & \ldots & 0 \\
 & 0 & p\ns_N & p\ns_{N-1} & p\ns_{N-2} & \ldots & p\ns_1 & p\ns_0 & 0 \ldots & 0 \\
 & \vdots\ &&&&&&&&\vdots\\
 & 0 & \ldots\ & 0 & p\ns_N & p\ns_{N-1} & p\ns_{N-2} & \ldots & p\ns_1 & p\ns_0 \\
 & N\,p\ns_N & (N-1)\,p\ns_{N-1} & (N-2)\,p\ns_{N-2} & \ldots\ & 1\,p\ns_1 & 0 & \ldots &\ldots & 0 \\
 & 0 & N\,p\ns_N & (N-1)\,p\ns_{N-1} & (N-2)\,p\ns_{N-2} & \ldots\ & 1\,p\ns_1 & 0 & \ldots & 0 \\
 & \vdots\ &&&&&&&&\vdots\\
 & 0 & 0 & \ldots & 0 & N\,p\ns_N & (N-1)\,p\ns_{N-1} & (N-2)\,p\ns_{N-2}& \ldots\ & 1\,p\ns_1 \\
\end{matrix}\right]
\label{discriminantfull}
\end{equation}
\end{widetext}
Since $p\ns_l$ is of maximal degree $l\Lambda\ns_\mu$ in $z\ns_\mu$, the discriminant as shown above must be of maximal degree
\begin{equation}
\textsf{deg}\,D(z) = 2\Lambda\ns_\mu(1+2+\ldots+N)=N(N-1)\Lambda\ns_\mu\ .
\end{equation}
More generally, the resultant of two polynomials disappears whenever the two polynomials have a common root. 

\subsection{Alternatives to the discriminant}
\label{app:alt}
When $N>2$, the roots of the discriminant gives us all the possible branch points, even those not associated with the $m^{\rm th}$ energy sheet that we desire to be
almost flat. Consequently, the flatness of the desired band in $\Htil(\Bk)$ may be underestimated. To remedy this, we may alternatively {\it define} $g\ns_\mu$ to include only
the roots of $(\ve\ns_m-\ve\ns_{m+1})^2(\ve\ns_m-\ve\ns_{m-1})^2$. However, this procedure may be more complicated to perform analytically, involving the explicit
solution of the degree $N$ characteristic polynomial. An analytic solution may not even exist for $N\geq 5$ due to the Abel-Ruffini Theorem~\cite{lang2013},
although this is not too constraining since most interesting flat band models in the literature contain no more than $N=4$ bands.

\end{document}